\documentclass[journal, onecolumn]{IEEEtran}
\usepackage{graphicx}
\usepackage{amsmath}

\begin{document}

\title{TPC Together with Overlapped Time Domain Multiplexing System Based on Turbo Structure}

\author{Hao~Zheng,
        ~Mingjun~Xing,
        ~Yutao~Yue,
        ~Xue~Li,
        ~Daoben~Li
        and Chunlin~Ji\\
{Kuang-Chi Institute of Advanced Technology, Shenzhen, China
 \\Email: \{hao.zheng, chunlin.ji\}@kuang-chi.org}
}


\maketitle

\begin{abstract}
Overlapped time domain multiplexing (OvTDM) is a novel technique for utilizing inter-symbol interference (ISI) to benefit a communication system. We implement the OvTDM technique based on turbo structure and associate a turbo product code (TPC) to construct a novel coded turbo-structure OvTDM system. Two schemes of the iterative receiver and soft input and soft output (SISO) decoding algorithms are presented. Simulation results show the advantage of structures in this paper. In addition, an attractive transmission rate and symbol efficiency of the designed system can also be observed.

\end{abstract}


\IEEEpeerreviewmaketitle

\section{Introduction}
It is well known that most traditional communication systems are designed based on Nyquist criterion \cite{Proakis12001}, in which inter-symbol interference (ISI) should be avoided between consequent symbols. In fact, the communication system without ISI is physically unrealizable. On the other hand, people tend to design a communication system with controlled ISI, such as Faster-than-Nyquist (FTN) signaling \cite{Mazo11975} and partial response signaling (PRS) \cite{Kabal11975}. However, these methods also treat the overlap between symbols as interference and do not really utilize it to collect the extra gain.\par

Based on ISI to benefit a communication system, overlapped time domain multiplexing (OvTDM) is proposed in \cite{Li12014}-\cite{Ji12016}. The idea of OvTDM is to shift a data-weighted and band-limited multiplexing waveform in the time domain to achieve an overlap between different transmitted symbols and a high transmission rate. It can help to form a convolution structure among consequent symbols, so OvTDM can also be regarded as one kind of waveform convolution coding. In the OvTDM system, these overlapped parts are never regarded as ISI but rather as a beneficial encoding constraint relationship. Therefore, OvTDM can show great performance of the transmission rate and the symbol efficiency that is defined as bits per symbol \cite{Li12014}\cite{Li22013}. Maximum likelihood sequence detection (MLSD) \cite{Forney11972} and maximum a posteriori (MAP) detection can be utilized to detect OvTDM signals. From the point of view of waveform ��convolution�� coding, the detection process can also be called OvTDM decoding.\par

Most previous studies have focused on the single structure of the OvTDM system. However, another way to improve the OvTDM system is to extend the coding structure \cite{Dong12013}\cite{Liu12013}. So, we construct a turbo structure inspired by the turbo code \cite{Berrou11996} for OvTDM. In addition, turbo product code (TPC) is employed as the forward-error-correction (FEC) module together with the turbo structure of OvTDM. In comparison to another popular FEC code, the low-density parity-check code (LDPC) \cite{Gallager11962}\cite{MacKay11996}, TPC is suitable to be constructed with a shorter code length and requires fewer iterations for decoding \cite{Pyndiah11998}\cite{Fang12000}. Finally, a coding system with three layers is formed, which contains the FEC code, the turbo structure and OvTDM respectively. Comparative simulation studies with the coded QAM system show the significant advantage of the coded turbo-strucutre of OvTDM.\par

\section{System Description}

\subsection{OvTDM Scheme}
In the OvTDM system, we artificially introduce ISI to form an overlap among different symbols. The mapping relationship between original bits and constellation symbols can follow the rule of ordinary modulation methods. Assuming the transmitted signals followed BPSK as ${\bf{x}} = [{x_0},{x_1}, \cdots ,{x_{L-1}}]$ with length $L$ and the multiplexing waveform as $h(t),t \in [0,{T_s})$ with symbol duration $T_s$, then the transmitted signal after overlapping can be expressed as
\begin{equation}
\label{eq1}
s(t) = \sum\limits_{i = 0}^{L - 1} {{x_i}h(t - i{T_s}/K)} = \sum\limits_{i = 0}^{L - 1} {{x_i}h(t - i\Delta T)}
\end{equation}
where $\Delta T = {T_s}/K$ is the time shift between symbols. In (\ref{eq1}), $K$ is the number of overlapped symbols during $\Delta T$, which is named the overlapping coefficient or the constraint length. Notice that, the larger the coefficient $K$ is, the more serious the ISI introduced. \par

\subsection{Turbo-Structure OvTDM with FEC}
\begin{figure}[!t]
\centering
\includegraphics[width=2.5in,height=1in]{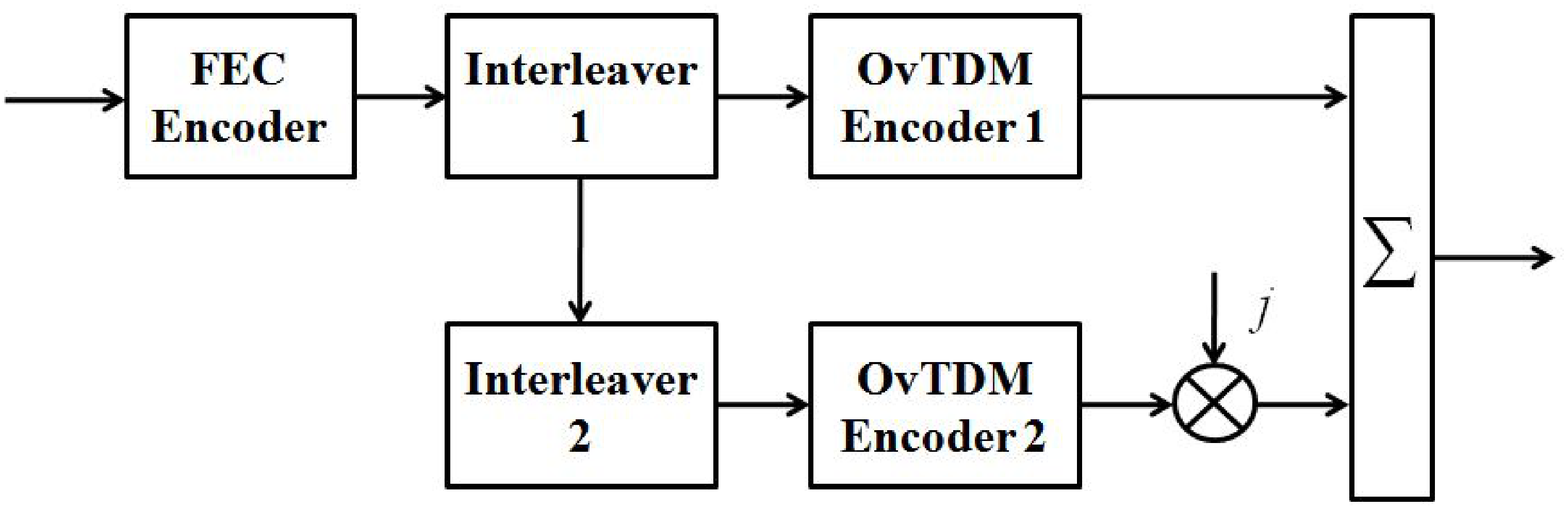}
\caption{The transmitter structure of the turbo-structure OvTDM together with TPC.}
\label{Fig_Transmitter}
\end{figure}

Fig.\ref{Fig_Transmitter} shows the transmitter of the turbo-structure OvTDM with FEC. The coded sequence that has passed the FEC encoder and one interleaver is sent to the first OvTDM encoder for the \emph{I} channel. Meanwhile, the same sequence is sent to pass the other interleaver to form another sequence with a different order, which is encoded by the second OvTDM encoder for the \emph{Q} channel. The output sequences from both two OvTDM encoders are combined to form a complex sequence. \par
The decoding process at the receiver is the key to the system design. It is based on the idea of iteration and the extrinsic information transformation \cite{Berrou11996}. During each iteration, the extrinsic information is exchanged between different decoders. The interleaver and the de-interleaver are employed to match the order of received sequences. Exchanging extrinsic information with low correlation can help to improve performance with the increase of iterations. Together with FEC, two schemes are addressed as follows:\par
\begin{figure}[!t]
\centering
\includegraphics[width=2.5in,height=1.2in]{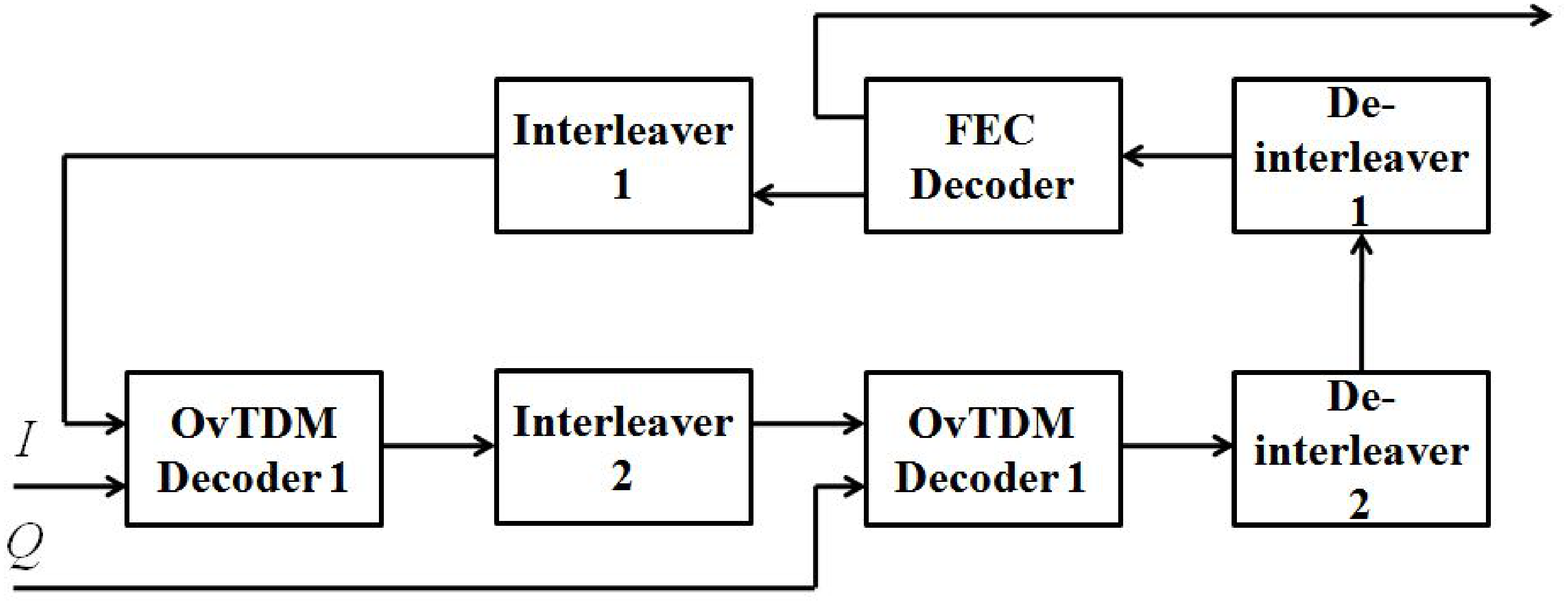}
\caption{The receiver structure of the turbo-structure OvTDM together with TPC (Scheme A).}
\label{Fig_Receiver_A}
\end{figure}

\emph{Scheme A:} After one round of decoding between two OvTDM decoders, the soft information is sent to the FEC decoder. Then the FEC decoder sends the soft information back to the OvTDM decoder. The model is shown in Fig.\ref{Fig_Receiver_A}. In this scheme, the FEC decoder needs to be involved in every iteration of the turbo structure.\par
\begin{figure}[!t]
\centering
\includegraphics[width=2.5in,height=1.2in]{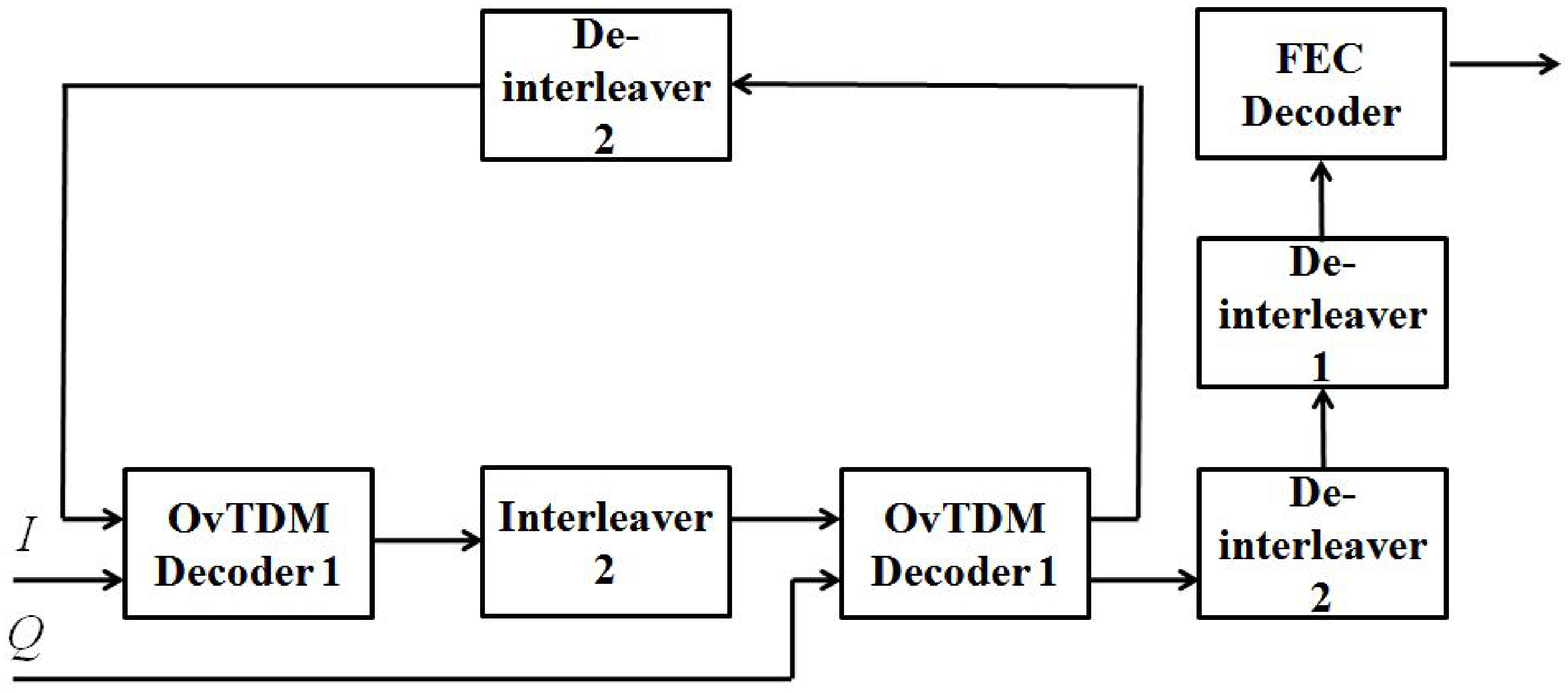}
\caption{The receiver structure of the turbo-structure OvTDM together with TPC (Scheme B).}
\label{Fig_Receiver_B}
\end{figure}

\emph{Scheme B:} As shown in Fig.\ref{Fig_Receiver_B},
OvTDM decoders work iteratively and do not exchange soft information with the FEC decoder at first. After several iterations, the soft information is sent to the FEC decoder. Unlike in Scheme A, the soft information is only exchanged once between OvTDM decoders and the FEC decoder.\par
The Viterbi algorithm \cite{Viterbi11967} is a good choice for detecting the OvTDM signals and selecting the possible sequence that is nearest to the received sequence \cite{Li12014}\cite{Li22013}. However, in the turbo structure, we need to exchange the extrinsic information for original bits, so soft output detecting algorithms \cite{Bahl11974}\cite{Ji22016} are more appropriate.\par

\section{Decoding Algorithms}

\subsection{MAP Algorithm for OvTDM}
As discussed in the above sections, OvTDM utilizes the ISI as the encoding constraint. Thus, it can also be represented as a trellis graph \cite{Li22013}. The BCJR algorithm \cite{Bahl11974} is regarded as an optimal MAP method based on the trellis graph, so it can be modified to calculate the maximum a posteriori probability (APP) for OvTDM encoded bits. \par

Denoting the input bit at time $t$ as $x_t$ and the received sequence at the receiver as ${\bf{r}}$ with length $N$, the log-likelihood-ratio (LLR) of APP of $x_t$ is
\begin{equation}
\label{eq3}
\begin{split}
{\lambda _t} &= \log \frac{{{\rm{p}}({x_t} =  + 1|{\bf{r}})}}{{{\rm{p}}({x_t} =  - 1|{\bf{r}})}} \\&= \log \frac{{\sum\limits_{\left( {{S_{t - 1}},{S_t}} \right),{x_t} =  + 1} {{\rm{p}}({S_{t - 1}},{S_t},{\bf{r}})} }}{{\sum\limits_{\left( {{S_{t - 1}},{S_t}} \right),{x_t} =  - 1} {{\rm{p}}({S_{t - 1}},{S_t},{\bf{r}})} }}
\end{split}
\end{equation}
where $S_t$ and $S_{t-1}$ are assumed as states of time $t$ and $t-1$ based on the trellis graph. According to properties of OvTDM, ${{\rm{p}}({S_{t - 1}},{S_t},{\bf{r}})}$ can be further expressed by
\begin{equation}
\label{eq3_1}
{\rm{p}}({S_{t - 1}},{S_t},{\bf{r}}) = {\alpha _t}({S_{t - 1}}){\gamma _t}({S_{t - 1}},{S_t}){\beta _t}({S_t})
\end{equation}
where ${\alpha _t}({S_{t - 1}})$ and ${\beta _t}({S_t})$ can be calculated by forward and backward recursion:
\begin{equation}
\label{eq14}
\begin{split}
{\alpha _t}({S_{t - 1}})
= \sum\limits_{{S_{t - 1}}} {{\alpha _{t - 1}}({S_{t - 2}})} {\gamma _t}({S_{t - 1}},{S_t})
\end{split}
\end{equation}
\begin{equation}
\label{eq15}
\begin{split}
{\beta _t}({S_t})
= \sum\limits_{{S_{t + 1}}} {{\beta _{t + 1}}({S_{t + 1}})} {\gamma _{t + 1}}({S_t},{S_{t + 1}})
\end{split}
\end{equation}
Assuming the corresponding output bit at time $t$ after OvTDM encoding as $y_t$, then
\begin{equation}
\label{eq13}
\begin{split}
{\gamma _t}({S_{t - 1}},{S_t})
= {\rm{p}}({x_t}){\rm{p}}({r_t}|{y_t})
\end{split}
\end{equation}

It is worth noting that there is no input bit in the tail part of OvTDM. So, ${\beta _L}({S_L})$ can be initialized directly. In the AWGN channel,
\begin{equation}
\label{eq17}
{\beta _L}({S_L}) = \frac{1}{{{{\left( {\sqrt {2\pi } \sigma } \right)}^K}}}\exp \left( { - \frac{{\sum\limits_{i = L}^N {{{({r_t} - {y_t})}^2}} }}{{2{\sigma ^2}}}} \right)
\end{equation}
where ${\sigma ^2}$ is the variance of noise.\par
Let LLR of a prior probability and the extrinsic information be ${\mu _t}$ and ${e_t}$, in the iterative decoder, the extrinsic information can be obtained by
\begin{equation}
\label{eq12}
{e_t} = {\lambda _t} - {\mu _t}
\end{equation}
which is the output of the OvTDM decoding module.

\subsection{FBBA for TPC}
One mainstream method of TPC decoding algorithm is augmented list decoding (ALD) \cite{Pyndiah11998}\cite{Fang12000}. The key idea of ALD is to form a list including the most likely codewords. Based on ALD, the Fang-Battail-Buda-Algorithm (FBBA) \cite{Fang12000} is an efficient SISO algorithm for TPC decoding to achieve near-optimum performance. FBBA is concluded as follows:\par

\emph{Step 1}: Sort the received symbols ${\bf{d}}$ in a decreasing order according to the LLR metric ${\bf{l}}$.\par
\emph{Step 2}: Permute the check matrix ${\bf{H}}$ according to the permutation pattern from the \emph{Step 1}. Then, it has to be adjusted by Gauss-Jordan elimination to obtain a systematic one ${{\bf{H}}^{\bf{\pi }}}$ that is used to re-code the component codeword to generate a new one ${{\bf{c}}^{\bf{\pi(0)}}}$.\par
\emph{Step 3}: A codebook list is obtained through the reversal of certain positions of ${{\bf{c}}^{\bf{\pi(0)}}}$ and sorted in an increasing order according to
\begin{equation}
\label{eq14}
Z({\bf{l}},{{\bf{c}}^{{\bf{\pi (i)}}}}) = - \sum\limits_{j = 0}^{n - 1} {\log \frac{{{\rm{p}}({l_j}|c_j^{^{\pi (i)}})}}{{{\rm{p}}({l_j}|c_j^{^{\pi (0)}})}}}
\end{equation}
where ${{\bf{c}}^{{\bf{\pi (i)}}}}$ and $n$ are denoted as the $i$th codeword in the codebook list and the component codeword length.\par
\emph{Step 4}:
The soft output can be calculated by the first codeword ${{\bf{c}}^{\bf{\pi(0)}}}$ and the opposite to the first codeword in $j$th position in the codebook list.\par
\begin{equation}
\label{eq15}
{\rho _j} = \frac{1}{4}(||{\bf{l}} - {{\bf{c}}^{{\bf{\pi (opp)}}}}|{|^2} - ||{\bf{l}} - {{\bf{c}}^{{\bf{\pi (}}0{\bf{)}}}}|{|^2})
\end{equation}
Following the above process, soft outputs can be calculated. Generally, four iterations are sufficient for the BER performance to converge.

\section{Simulation Studies}
In this section, we need to investigate the performance through some comparative simulation. We choose the Chebyshev waveform with attenuation level 80dB as the multiplexing waveform for OvTDM. Extended ${\rm{BCH}}(64,57)$ is employed as the component codeword to construct a squared TPC. Thus, the code rate is ${R_{TPC}}{\rm{ = (57/64}}{{\rm{)}}^2}{\rm{ = 0}}{\rm{.7932}}$. The AWGN channel is considered as the transmission channel in the simulation.\par
\begin{figure}[!t]
\centering
\includegraphics[width=3in, height=2in]{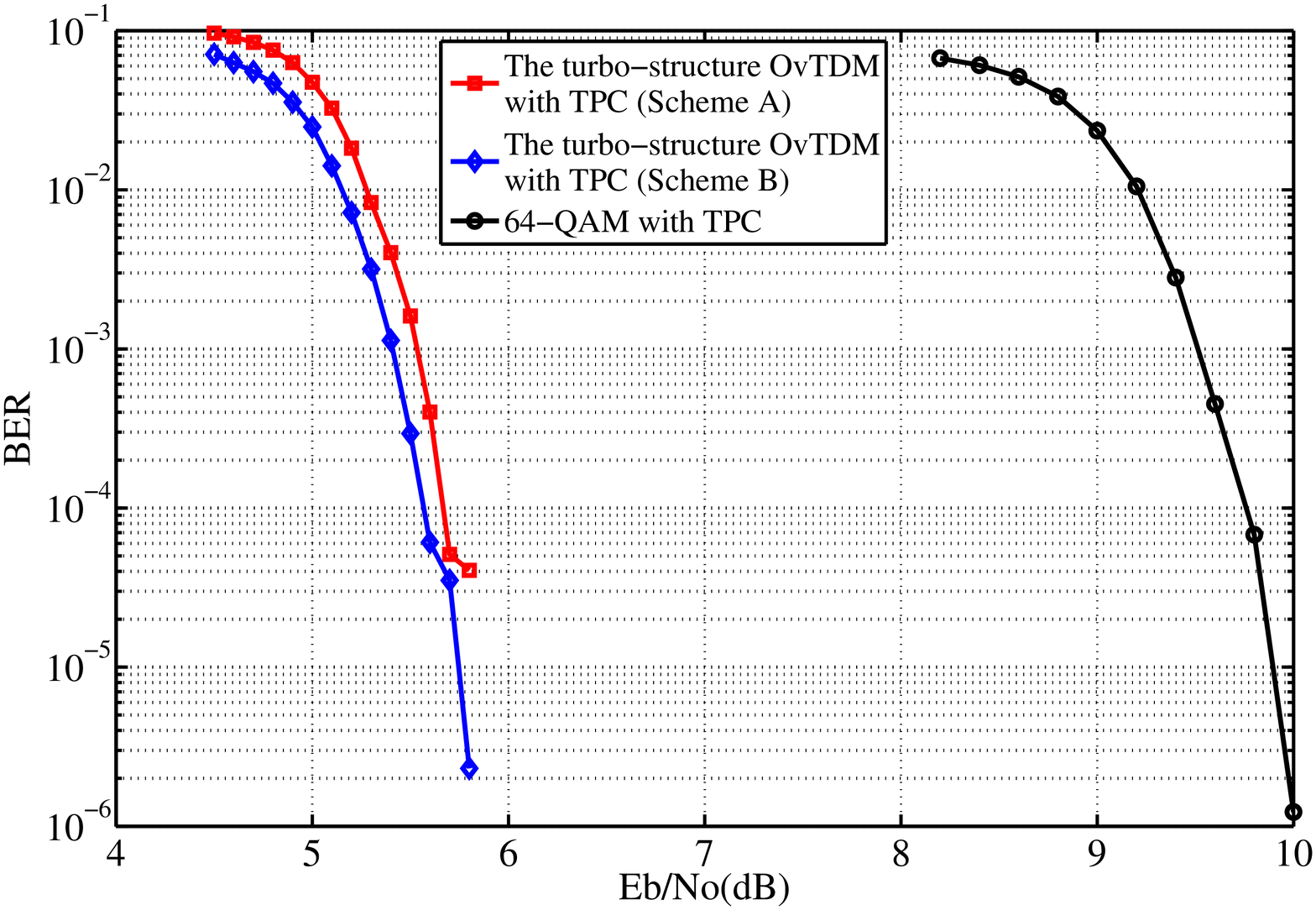}
\caption{BER performance of the turbo-structure OvTDM ($K=6$) and 64-QAM with TPC$(64,57)^2$.}
\label{Fig_K6_Comp}
\end{figure}
\begin{figure}[!t]
\centering
\includegraphics[width=3in, height=2in]{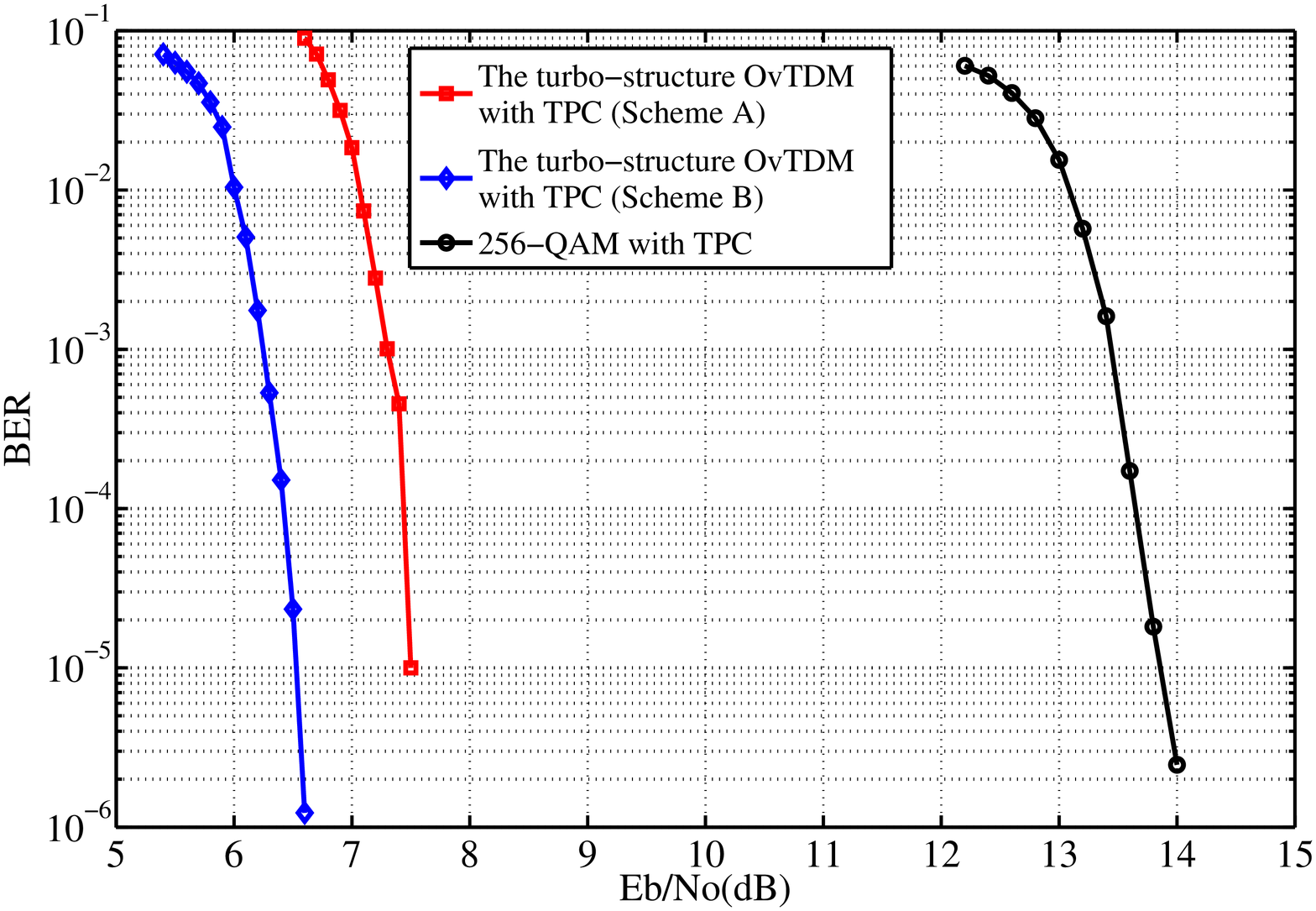}
\caption{BER performance of the turbo-structure OvTDM ($K=8$) and 256-QAM with TPC$(64,57)^2$.}
\label{Fig_K8_Comp}
\end{figure}
As mentioned before, the turbo-structure OvTDM uses the same information sequence for both \emph{I} and \emph{Q} channels, so its equivalent coding rate with TPC is ${R_{OvT}} = 1/2 \cdot {R_{TPC}} \cdot L/N$. When the length of the information sequence is large enough, $L/N \approx 1$, so we ignore it in the simulation. BPSK is used as the original modulation for the OvTDM system. Thus, the symbol efficiency of the turbo-structure OvTDM with TPC is ${\eta _{OvT}} = {R_{OvT}} \cdot 2K = {R_{TPC}} \cdot K$ (bits/symbol). On the other hand, if we select $M$-ary QAM with TPC for comparison, its symbol efficiency is ${\eta _{QAM}} = \log 2(M) \cdot {R_{TPC}}$ (bits/symbol). In order to do the comparative studies under the same symbol efficiency, we select $K=6$ and $64$-QAM in Fig.\ref{Fig_K6_Comp} as well as $K=8$ and $256$-QAM in Fig.\ref{Fig_K8_Comp}. The BER plots in both Fig.\ref{Fig_K6_Comp} and Fig.\ref{Fig_K8_Comp} show the significant advantage of the coded turbo-structure OvTDM. In Fig.\ref{Fig_K6_Comp}, the same symbol efficiency is $4.7592$ (bits/symbol) and the required ${E_b}/{N_0}$ of $64$-QAM with TPC to achieve the BER $ < {10^{ - 5}}$ is $10$dB, but the turbo-structure OvTDM with TPC can achieve BER $ < {10^{ - 5}}$ at $5.8$dB. In Fig.\ref{Fig_K8_Comp}, the same symbol efficiency is $6.3456$ (bits/symbol) and the required ${E_b}/{N_0}$ of the turbo-structure with TPC using Scheme B to achieve the BER $ < {10^{ - 5}}$ is $7.4$dB less than that of 256-QAM with TPC.\par
\begin{figure}[!t]
\centering
\includegraphics[width=3in, height=2in]{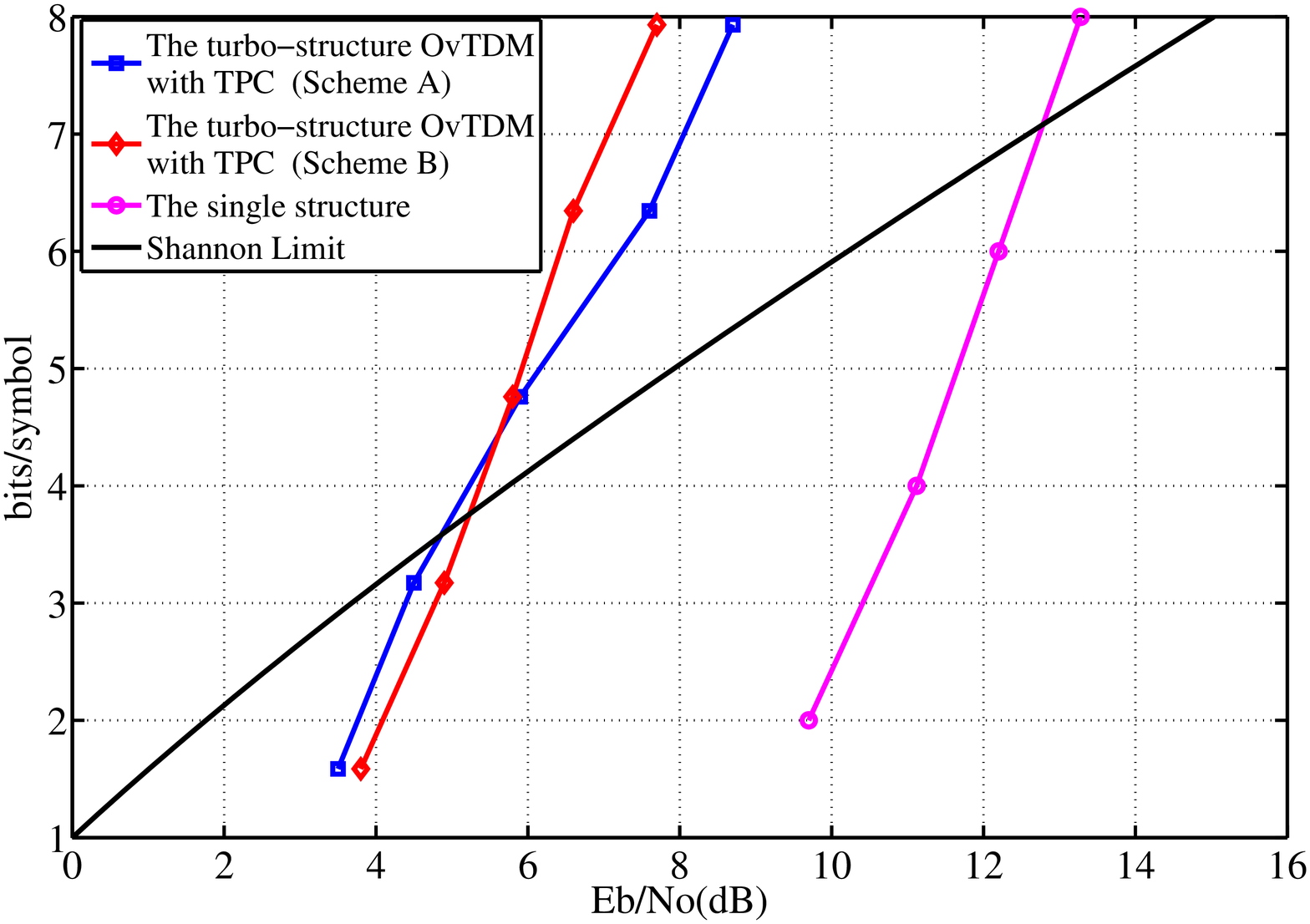}
\caption{Comparison of the symbol efficiency (bits/symbol) among different schemes.}
\label{Fig_Sym_Eff_Comp}
\end{figure}
Moreover, Fig.\ref{Fig_Sym_Eff_Comp} illustrates the comparison result of the symbol efficiency among different schemes when BER $ < {10^{ - 5}}$. The symbol efficiency of the single structure OvTDM has been shown in \cite{Li22013}. Also, we plot the corresponding symbol efficiency of the Shannon Limit \cite{Shannon11948}. In Fig.\ref{Fig_Sym_Eff_Comp}, the turbo-structure OvTDM with TPC has a obvious improvement, compared with the single structure OvTDM. On the other hand, with the increase of the symbol efficiency, schemes in this paper can achieve it at a lower ${E_b}/{N_0}$ than the Shannon Limit that represents traditional communication systems.\par

\section{Conclusion}
This paper mainly focuses on structures and SISO decoding algorithms of the turbo-structure OvTDM with TPC, which demostrate a significant improvement over the single structure OvTDM. In addition, compared with the coded QAM system of the same symbol efficiency, the BER performance of the turbo-structure OvTDM with TPC is much better. Simulation results shows the advantage of the turbo-structure OvTDM with TPC in the communication scenario requiring high transmission rate at a relatively low ${E_b}/{N_0}$.\par



\begin{thebibliography}{99}

\bibitem{Proakis12001}
J.~G.~Proakis, Digital Communications, 4th ed, Macgraw Hill, New York, 2001.

\bibitem{Mazo11975}
J.~E.~Mazo, ``Faster-than-nyquist signaling," Bell Syst. Tech. J., vol. 54, no. 8, pp. 1451-1462, Oct. 1975.

\bibitem{Kabal11975}
P.~Kabal and S.~Pasupathy, ``Partial-Response Signaling," IEEE Trans. Commun., vol. 23, no. 9, pp. 921-934, Sep. 1975.

\bibitem{Li12014}
D.~Li, ``A novel high spectral efficiency waveform coding-OVTDM," International Journal of Wireless Communications and Mobile
Computing, Special Issue: 5G Wireless Communication Systems, vol. 2, no. 4-1, pp. 11-26, Dec. 2014.

\bibitem{Li22013}
D.~Li, Waveform Coding Theory of High Spectral Efficiency--OVTDM and Its Application, Science Press, Beijing, 2013.

\bibitem{Ji12016}
C.~Ji and R.~Liu, ``Study on a High Spectrum Modulation by Introducing Intersymbol Interference," in Proc. 2016 IEEE Int. Conf. on Signal Processing, Communications and Computing, Hong Kong, , pp. 2904-2909, Aug. 2016.

\bibitem{Forney11972}
G.~D.~Forney, ``Maximum Likelihood Sequence Estimation of Digital Sequences in the presence of intersymbol interference," IEEE Trans. Inform. Theory, vol 18, no. 3, pp. 363-378, May 1972.

\bibitem{Dong12013}
X.~Dong, ``Research on the performance of OvTDM and Turbo-OvTDM technology application in multi-carrier system," M.Sc. Dissertation, Beijing Univ. Posts Telecommun., 2013.

\bibitem{Liu12013}
B.~Liu, ``Applications of overlapped multiplexing principle in telecommunications," Ph.D. Dissertation, Beijing Univ. Posts Telecommun., 2014.

\bibitem{Berrou11996}
C.~Berrou and A.~Glavieux, ``Near optimum error correcting coding and
decoding: Turbo codes," IEEE Trans. Commun., vol. 44, no. 10, pp. 1261-1271,
Oct. 1996.

\bibitem{Gallager11962}
R.~Gallager, ``Low-density parity-check codes," IRE Trans. Inform. Theory, vol. 8, no. 1, pp. 21-28, Jan. 1968.

\bibitem{MacKay11996}
D. J. C. MacKay and R. M. Neal, ``Near shannon limit performance of low density parity check codes," Electron. Lett., vol. 33, no. 6, pp. 457-458, Mar. 1997.

\bibitem{Pyndiah11998}
R.~Pyndiah, ``Near optimum decoding of product codes: block turbo codes," IEEE Trans. Commun., vol. 46, no. 8, pp. 1003-1010, Aug. 1998.

\bibitem{Fang12000}
J.~Fang, F.~Buda and E.~Lemois, ``Turbo Product Code: a well suitable solution to wireless packet transmission for very low error rates," in Proc. 2nd Int. Symp. on Turbo
Codes \& Related Topics, France, pp. 101-111, Sep. 2000.

\bibitem{Viterbi11967}
A.~J.~Viterbi, ``Error bounds for convolutional codes and an asymtotically
optimum decoding algorithm," IEEE Trans. Inform. Theory, vol. 13, no. 2, pp. 260-269, Apr. 1967.

\bibitem{Bahl11974}
L.~R.~Bahl, J.~Cocke, F.~Jelinek, and J.~Raviv, ``Optimal decoding of linear
codes for minimizing symbol error rate," IEEE Trans. Inform. Theory, vol. 20, no. 2,  pp. 284-287, Mar. 1974.

\bibitem{Ji22016}
C.~Ji, ``On Sequential Learning for Parameter Estimation in Particle Algorithms for State-Space Models," International Journal of Statistics and Probability, vol. 6, no. 1, pp. 13-23, Jan. 2017.

\bibitem{Shannon11948}
C.~E.~Shannon, ``A mathematical theory of communication," Bell Syst. Tech. J., vol. 27, no. 7, pp. 379-423, 1948.
\end{thebibliography}

\end{document}